\documentclass{aastex62}

\DeclareFixedFont{\ttb}{T1}{txtt}{bx}{n}{12} 
\DeclareFixedFont{\ttm}{T1}{txtt}{m}{n}{12}  

\usepackage{color}
\definecolor{deepblue}{rgb}{0,0,0.5}
\definecolor{deepred}{rgb}{0.6,0,0}
\definecolor{deepgreen}{rgb}{0,0.5,0}

\usepackage{listings}

\newcommand\pythonstyle{\lstset{
language=Python,
basicstyle=\ttm,
otherkeywords={self},             
keywordstyle=\ttb\color{deepblue},
emph={MyClass,__init__},          
emphstyle=\ttb\color{deepred},    
stringstyle=\color{deepgreen},
frame=tb,                         
showstringspaces=false            %
}}

\lstnewenvironment{python}[1][]
{
\pythonstyle
\lstset{#1}
}
{}

\newcommand\pythoninline[1]{{\pythonstyle\lstinline!#1!}}

\graphicspath{{./}{figures/}}

\submitjournal{ApJ}

\shorttitle{Analysis of the secondary clump red giant HD~226808}
\shortauthors{De Moura et al.}

\begin{document}

\title{Spectroscopic and asteroseismic analysis of the secondary clump red giant HD~226808\footnote{KIC~5307747}.}

\correspondingauthor{J.D.N.Jr.}
\email{jdonascimento@fisica.ufrn.br}

\author{Bruno Lustosa De Moura}
\affil{Departamento de Fisica, Universidade Federal do Rio Grande do Norte, 59072-970 Natal, RN, Brazil}
\affil{Instituto Federal do Rio Grande do Norte - IFRN, Brazil}

\author{Paul G. Beck}
\affiliation{Instituto de Astrof\'{\i}sica de Canarias, E-38200 La Laguna, Tenerife, Spain}
\affiliation{Institute of Physics, Universit\"at Graz, NAWI Graz, Universitaetsplatz 5/II, 8010 Graz, Austria}
\affil{Departamento de Astrof\'{\i}sica, Universidad de La Laguna, E-38206 La Laguna, Tenerife, Spain}

\author{Leandro de Almeida}
\affil{Departamento de Fisica, Universidade Federal do Rio Grande do Norte, 59072-970 Natal, RN, Brazil}

\author{Tharcisyo S. S. Duarte}
\affil{Instituto de Forma\c{c}\~ao de Educadores, Universidade Federal do Cariri - UFCA, Brazil}

\author{Hugo R. Coelho}
\affil{Departamento de Fisica, Universidade Federal do Rio Grande do Norte, 59072-970 Natal, RN, Brazil}

\author{Jefferson S. da Costa}
\affil{ECT, Universidade Federal do Rio Grande do Norte, 59072-970 Natal, RN, Brazil}

\author{Matthieu Castro}
\affil{Departamento de Fisica, Universidade Federal do Rio Grande do Norte, 59072-970 Natal, RN, Brazil}

\author{Jos\'e-Dias do Nascimento Jr.}
\affil{Departamento de Fisica, Universidade Federal do Rio Grande do Norte, 59072-970 Natal, RN, Brazil}
\affil{Harvard-Smithsonian Center for Astrophysics, 60 Garden St., Cambridge, MA 02138}



\begin{abstract}
	In order to clarify the properties of the secondary clump star HD 226808 (KIC 5307747),
	we combined four years of data from the \textit{Kepler} space
	photometry with the high-resolution spectroscopy of the HERMES spectrograph mounted on the Mercator telescope. The fundamental 
	atmospheric parameters, radial velocities, rotation velocities and elemental abundance for Fe and Li were determined by analyzing line 
	strengths and fitting line profiles, based on a 1D LTE model atmosphere. Secondly, we analyzed photometric light curve obtained by \textit{Kepler} 
	and we extracted asteroseismic data  of this target by using LAURA (Lets Analysis, Use and Report of Asteroseismology), 
	a new seismic tool developed for the study of evolved  FGK solar-like stars.  We determined the evolutionary status and  
	effective temperature; surface gravity;  metallicity,  microturbulence and chemical abundances for Li, Ti, Fe, and Ni  
	for   HD~226808,  by employing spectroscopy, asteroseismic scaling relations  and evolutionary structure models built in order to match  
	observed data.  Our results also show that an accurate  synergy between good spectroscopic analysis and asteroseismology can
	provide a jump toward  understanding evolved stars.
\end{abstract}



\section{Introduction} 
Red giants are cool evolved stars with an extended convective envelope, which can, as in main sequence solar-like stars, stochastically excite modes of oscillation. 
This stellar evolutionary phase is  well studied from both the observational and the theoretical aspects, but still quite intriguing since, after the exhaustion of 
H in the core, several structural changes occur in the interior of the star along this phase. In fact, the evolution proceed by fusion of hydrogen in a shell surrounding 
an inert core of helium,  which continues to increase its mass and contracts,  while the stellar radius expands.  If the stellar core reaches a temperature of 
$\sim$ 100 million K, the fusion of helium through the triple alpha process can start. However, the initial stellar mass will play a defining role on the evolution of the 
star after helium begins to burn in the core.  Stars with masses below $\sim 1.9 M_{\odot}$ develop degenerate helium cores which increase their mass while the H 
is burning in the shell,  until the core reaches a critical mass of $M_{c} \sim 0.47M_{\odot}$. 

\begin{figure*}
	\includegraphics[width=1.\columnwidth]{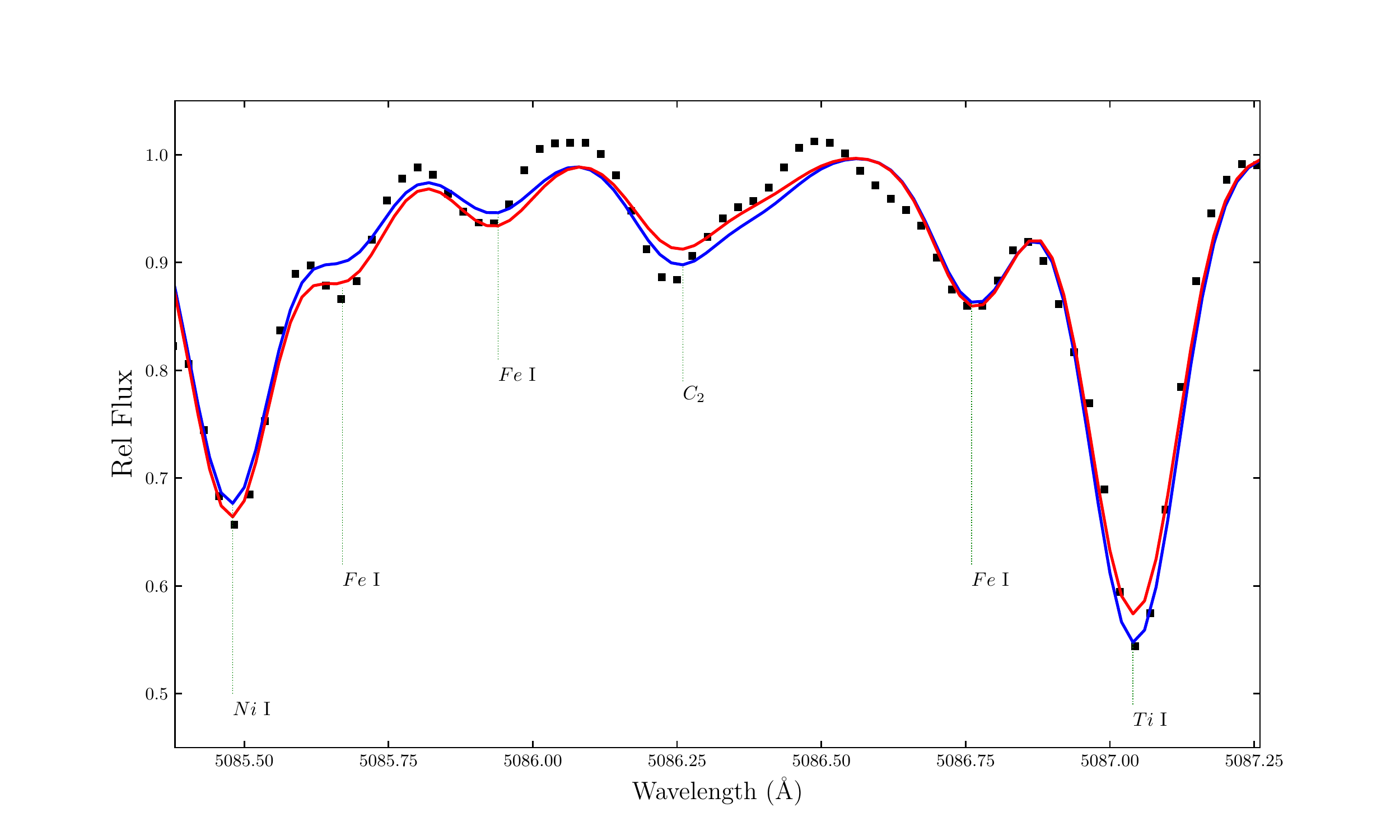}
	\caption{Observed and synthetics spectra around $C_2$ region around $\lambda$ 5086 A of HD~226808 (KIC~5307747) as observed by the MERCATOR telescopes. The observed spectrum is represented by black dots. \textbf{The synthetic spectra represented by red and blue lines  were computed with the same spectroscopic  fundamental parameters, $T_{eff} = 5065$, $log(g) = 2.99$, $[Fe/H] = 0.08$ and $V_t = 1.27$ and presented on the table 2.  The synthetic spectrum represented by the red line used $[C/H] = -0,05$; $[Ti/H] = -0,10$; $[Fe/H] = 0,01$; $[Ni/H] = 0,33$. The synthetic spectrum represented by the blue line used  $[C/H] = -0,08$; $[Ti/H] = -0,13$; $[Fe/H] = 0,08$; $[Ni/H] = 0,36$}. The comparison of synthetic and observed spectra results a 
        "o-c" (difference between the observed and synthetic spectra) of  about 0.02 dex for these species.}	
	\label{fig:spectra}
\end{figure*}

At this point,  a helium flash occurs abruptly  in the center of the helium core and the degeneracy is gradually lifted.  Those stars, due to the degenerate condition of the core,  
have similar core masses, hence similar luminosities, and hence are found in a narrow region of a H-R diagram, forming the so-called `red clump'  (RC)  
\citep[e.g.,][]{cannon1970,faulkner1973, girardi1998, girardi1999}. On the other hand, stars with masses higher than  $\sim1.9 M_{\odot}$ will not develop a degenerate 
helium core in the post-main sequence phase. They will not have an explosive liberation of energy like the one seen during the helium flash. Instead, they will have a gradual 
ignition of helium. Therefore, those stars will occupy a region of the H-R diagram similar to the red clump, but at lower luminosities and spread wide. This group, 
which includes  more massive core helium burning stars, at ages of about 1~Gyr, is referred as the secondary clump and its components are the 
second clump stars \citep[see e.g.,][]{girardi1999}. 

\cite{hawkins2017} studied in details the spectroscopic signatures of genuine core helium burning red clump giant stars and how they compare to shell hydrogen burning 
RGB stars. Additionally, spectroscopic study of genuine RGB or RC stars have been done by \cite{Anders2016} for some APOGEE and CoRoT targets. Red clump stars are natural 
standard candles \citep{stanek1998,hawkins2017}, while regular RGB stars or more massive secondary RC stars of nearly the same effective temperatures ($T_{eff}$) are not. In this 
context, finding  and characterizing  core helium burning RC stars  is very important to fine tune stellar evolution and Galactic archeology,  and consequently for building a  precise  
cosmic distance ladder \citep{stanek1998, bressan2013, bovy2014, gontcharov2017,hawkins2017}.  Distinguish RC stars from less  evolved shell hydrogen burning RGB stars or 
more massive secondary RC stars is a critical point and a bottle neck to solving some stellar astrophysics problems \citep{bressan2013}. 

\vspace{.3cm}

Fortunately, continuous photometric observations from space missions such as CoRoT \citep{baglin} and \textit{Kepler}   \citep{borucki2009}, 
provide photometric data of an unprecedented quality. This gives us the possibility to better distinguish the evolutionary status 
of red giants along  {the H-shell burning phase} before He-ignition from those on the He-burning phase after He-ignition, as described by \cite{bedding2011}. In particular, by using asteroseismic scaling
relations calibrated on solar values, it has been possible to determine with good accuracy stellar mass ($M$)  and radius ($R$)  \citep[e.g.,][]{pinsonneault2014, 
	casagrande2014} of hundreds of observed red-giant stars. Several progresses have been made during the last decade and asteroseismology provides a way 
forward to classify RC based on stellar natural oscillation frequencies. However at present, some RC stars, classified so far as authentic secondary 
clump giant stars \citep{mosser2014}, have a high resolution spectroscopic study. Therefore the importance of this analysis.  

In this study, we use photometric data from  \textit{Kepler}  space telescope and HERMES ground-based high-resolution spectroscopy to produce a  deep analysis of the bright red-giant star HD~226808 (KIC~530774). In particular, this star is one of three brightest classified as secondary clump star and observed by \textit{Kepler}   on long cadence mode \citep{mosser2014}. This paper is organized as follows: in 
Section \ref{sec:spectroscopy}, we present the spectroscopic observations and spectroscopic derived fundamental parameters.  
In Section~3 we present asteroseismic analysis. In Section~4 we compare the asteroseismic results with evolutionary models.  
Conclusions are presented in Section~5.

\section{Spectroscopic observations}  \label{sec:spectroscopy}

Our target, the star HD~226808 (KIC\,530774), with V $=$ 8.67 mag was observed on 2015 July 3 with the 
{\it High Efficiency and Resolution Mercator   \'Echelle Spectrograph} (HERMES, \citealt{Raskin2011,RaskinPhD}) 
mounted on the 1.2 m Mercator Telescope at the Observatorio del Roque  de los Muchachos on La Palma, Canary 
Islands, Spain. The HERMES spectra covers a wavelength range between 375  and 900\,nm with a spectral resolution 
of R$\simeq$85\,000. The  wavelength reference was obtained from emission spectra of  Thorium-Argon-Neon 
reference frames in close proximity to the individual  exposure. The standard steps of the spectral reduction
were performed with an instrument-specific pipeline as described by \cite{Raskin2011} and \cite{RaskinPhD}.  The 
radial velocity (RV) for each  individual spectrum was determined from the cross correlation of the stellar spectrum in 
the wavelength range between 478 and  653 nm with a standard mask optimized for Arcturus provided by the 
HERMES pipeline toolbox.  For HERMES, the 3$\sigma$ level of the night-to-night 
stability for the observing mode described above is $\sim$300\,m/s,  which is used as the classical threshold for RV variations 
to detect binarity. We corrected individual spectra for the Doppler shift before normalization and combined  individual spectra 
as described in \cite{beck2016}. The total integration time was 1.7~hrs,  split into four equal parts of 1500 seconds each. 
The final spectrum, stacked from an individual exposures shows S/N\,$\simeq$\,150 around the  Li\,\textsc{i}  $\lambda$ 670 nm line.

\subsection{Fundamental parameters from HERMES  spectroscopy}
\label{sec:spec_par}

For the fundamental parameters, the analysis of HD~226808  starts by using as a first guess  fundamental parameters (effective temperature  $T_{\rm eff}$; surface gravity 
$\log g$;  metallicity [Fe/H] and microturbulence)   from the revised   {\it Kepler}  Input   Catalog (KIC)   by  \cite{Huber2014}.  
Next,  we used the  ARES code (v2)  \citep{sousa2007,  sousa2015} to measure equivalent widths (EW) of selected spectra absorption  
Fe I and Fe II lines,   and with the q2 code  \cite{ramirez2014}  we determined  the fundamental physical parameters  based  on a 
 line list,  as described by  \cite{ramirez2014}.   Then,  we refine the  process  based   on  one-dimensional (1D) local thermodynamic equilibrium  (LTE) 
 model  atmosphere as described by  \cite{beck2016}.       We  also   considered    the  Arcturus  and  solar parameters   as   other   reference values,  as  described   by   	
\cite{hinkle2005},  \cite{ramirez2011}, \cite{mele2012},   \cite{monroe2013}, and  \cite{ramirez2014}.  Final spectroscopic parameters,  
such as the values of effective temperature $T_{\rm eff}$,  surface gravity  $\log g$, metallicity $[{\rm Fe/H}] $, and  A(Li) of HD~226808 are given  
in Table \ref{tab:01}.  Typically, stellar parameters  uncertainties were computed as descried by \cite{epstein2010} and \cite{bensby2014} 
 and are as  $^{+39}_{-29}$, $\pm0.08$, $\pm0.04$ and $\pm0.04$ for Teff, log g, [Fe/H], and $V_{\rm t}$, respectively.  
For abundances determinations, we used spectral synthesis based on  a  combination of a 2014 version of the  code  MOOG \citep{sneden1973}   with Kurucz atmosphere models  \citep{castelli2004} and a line lists as  described  in the  Table \ref{linelist}.  \textbf{(for fundamental parameters  determinations, we  used  a line list as described in the appendix \ref{appendixb})}.  A low lithium abundance signature is found for this secondary red clump giant star. However, on the same evolutionary stage, some lithium-rich stars has been found by \cite{kumar2018}. 
The values that we obtained for effective temperature and metallicity are comparable with those by \citet{takeda2015} who observed this star with the High Dispersion Spectrograph at the  Subaru telescope. The value of $\log g$ results to be slightly higher than the value obtained by \citet{takeda2015}, however within the error bar. In Figure \ref{fig:spectra} we show part of the  spectrum for HD 226808 in the region around 5086 \AA.  We also show results for two different synthetic spectra  computed with the code MOOG for the same spectroscopic  fundamental parameters. \textbf{For the error analysis in the abundances determination we preserve the same fundamental parameters and slightly vary the abundances along a step size from 0.05 until 0.10 dex. The values are shown in the table.}

 \begin{table}
        \caption{Line list  used for HD~226808 (KIC\,530774).}
        \centering
        \begin{tabular}{ccccc}
                \hline
                \multicolumn{1}{l}{$wavelength$} & \multicolumn{1}{c}{$species$} & \multicolumn{1}{c}{$\chi_{exc} $} & \multicolumn{1}{c}{$log (gf)$} & \multicolumn{1}{c}{\textbf{EW}}   \\
                \multicolumn{1}{c}{$\mbox{\normalfont\AA} $} & \multicolumn{1}{c}{$   $} & \multicolumn{1}{c}{$eV$} & \multicolumn{1}{c}{$    $} & \multicolumn{1}{c}{\textbf{$\mbox{\normalfont\AA} $}}   \\ \hline

 5085.477  &  28.0    &  3.658  &  -1.541 & \textbf{94.12}\\
 5085.676  &  26.0    & 4.178   & -2.610 & \textbf{9.78} \\
 5085.837  & 26.0     & 4.495   & -3.881 & \textbf{9.27} \\
 5085.933   &  26.0   &  3.943   & -3.151 & \textbf{9.77} \\
 5086.231  &  606.0  & 0.252   & -0.110 & \textbf{24.16} \\
 5086.251  &  26.0    & 4.988   & -2.624 & \textbf{24.16}  \\
 5086.259  &  606.0  & 0.252   & -0.122 & \textbf{24.16}   \\
 5086.390  &  606.0  & 0.252   & -0.133 & \textbf{---}   \\
 5086.765  &  25.0    & 4.435   & -1.324 & \textbf{36.07}     \\
 5086.805  &  606.0  & 0.301   & -0.649 & \textbf{36.07}      \\
 5086.997  &  10108.0 &0.040  &  -10.286  & \textbf{75.40}  \\
 5087.058  &  22.0       &1.430   & -0.780  & \textbf{75.43}    \\
 5087.065   & 606.0     &   0.301 &   -0.684 & \textbf{75.43}   \\
                \hline
        \end{tabular}
        \label{linelist}
        {\footnotesize \\
                \vspace{1mm}$^{*}$  For species  we are using MOOG standard  \\ notation for atomic or molecular identification. \\
                For example, 26 represent Fe(26) while the 0 after  \\ the  decimal indicates neutral and 1 singly ionized.  \\
                 $\chi_{exc} $ is the line excitation potential  in electron \\volts (eV)  }
\end{table}

\begin{table}
	\caption{Fundamental parameters data of HD~226808 (KIC\,530774).}
	\centering
	\begin{tabular}{lr}
		\hline
		Right Ascension (J2000.0)    & $19^{\rm h} 57^{\rm m} 34.0^{\rm s}$ \\
		Declination  (J2000.0)       & $+40\degr 37\arcmin 16.1\arcsec$\\
		Kp (mag)${}^{*}$       & 8.397 \\
		\hline
		\multicolumn{2}{c}{spectroscopically derived parameters} \\  
		\hline
		\ Teff (K)    & 5065 $^{+39}_{-29}$ \\
		log $g$ (cgs)        & $2.99 \pm 0.08$  \\
		  $V_{\rm t}$ (km~s$^{-1}$)         & $1.27 \pm 0.04$   \\
		$[{\rm Fe/H}]{}^{**}$   &   0.08$~\pm~$\textbf{0.04}$~dex$  \\
		$A(Li)$      			&  	 \textbf{$\leq$0.58}  \\ 
		$  A(C)  $    			&  	 {8.35~$\pm~$\textbf{0.05}$~ dex$}  \\ 
		$  A(Ti)  $ 			&   { 4.82~$\pm~$\textbf{0.06}$~dex$}  \\ 
		$ A(Fe)  $ 			&    { 7.58~ $\pm~$\textbf{0.07}$~dex$ } \\
		$ A(Ni)  $  		&   	 {6.58~$\pm~$\textbf{0.05}$~ dex$}  \\
				\hline
			\end{tabular}
	{\footnotesize \\
		\vspace{1mm}$^{*}$ Kp is the white light $\it Kepler$ magnitude taken from the \\
		revised {\it Kepler}  Input Catalog (KIC) \citep{Huber2014}. \\ $^{**}$ Abundances assuming  \citet{2009ARA&A..47..481A} }
	\label{tab:01}
\end{table}

\section{Analysis of asteroseismic data}

\begin{figure}
	\includegraphics[width=1.\columnwidth]{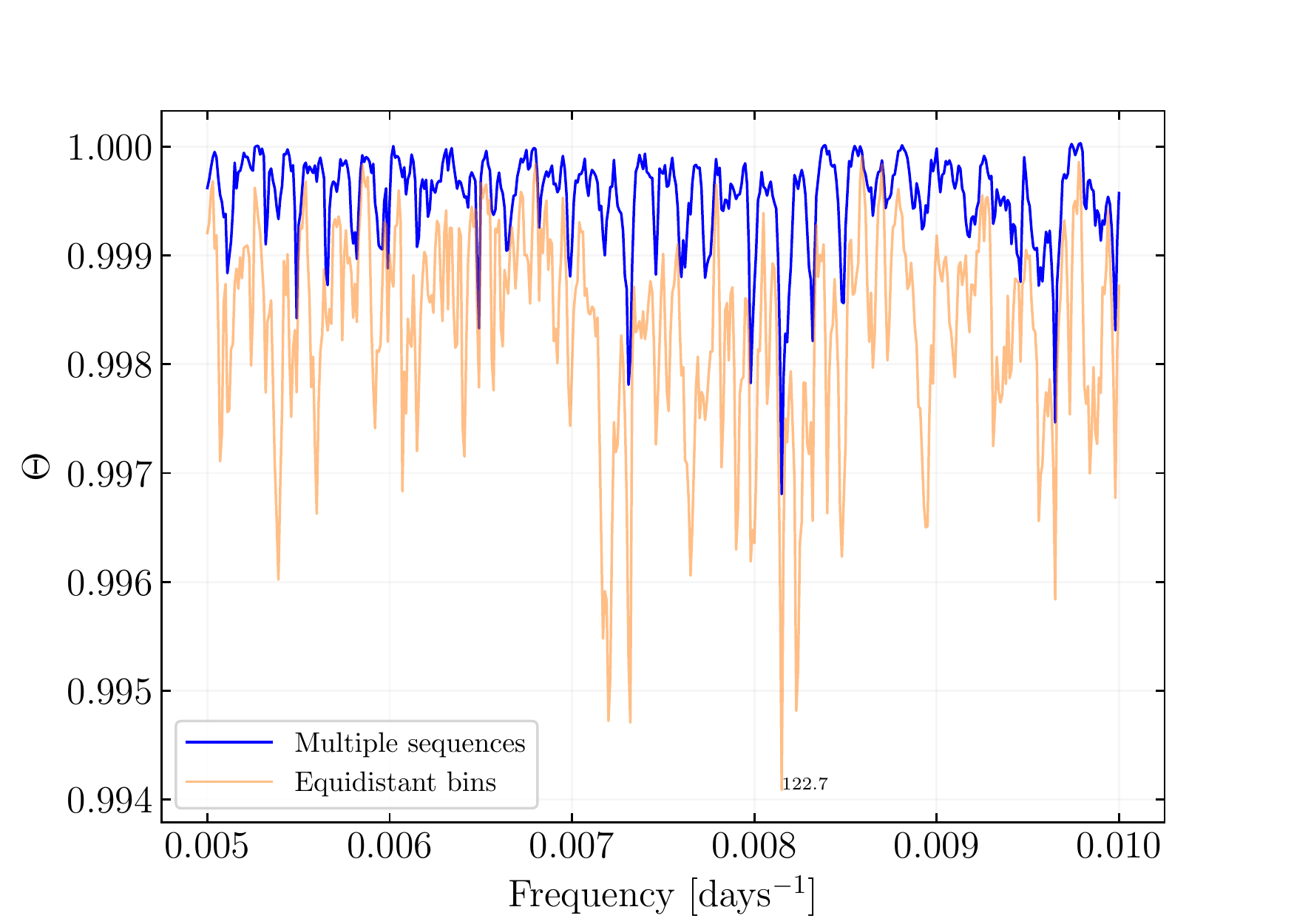}
	\caption{Determination of the period of rotation of HD~226808 (KIC~5307747) by phase dispersion minimization. The phase is modulated in the minimization of the phase graph by frequency. The smallest phase, for this modulation occurs around 0.008 which represents about 122 days. The check is made by general variance and equidistant or by multiple sequences represented by the two colors.}
	\label{pdm}
\end{figure}

\begin{figure}
	\includegraphics[width=1.\columnwidth]{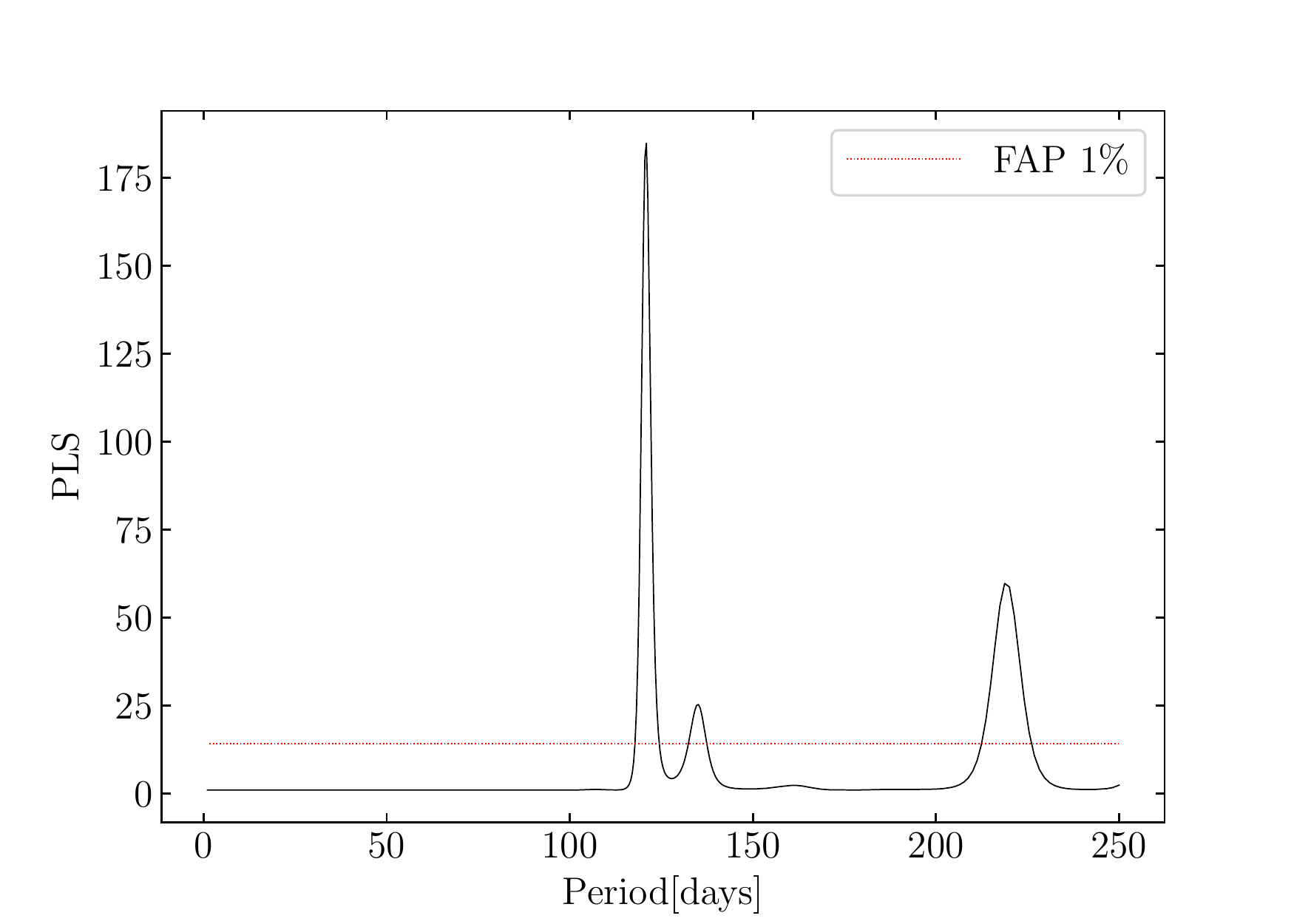}
	\caption{Lomb-Scargle periodogram of HD~226808 (KIC~5307747). The red line represents the 1\% probability limit of the signal.
		The highest peak indicates rotation. 
		}
	\label{ls}
\end{figure}

We have developed a code with the purpose of analyzing stellar oscillations and extracting seismic data from light curves. The code makes use of the version 3.6 of the python programming language and has been calibrated by using several stars from the literature, especially the giant stars from \cite{ceillier2017} and the KIC 4448777 from \citet{dimauro2018}. The \textit{Lets Analysis, Use and Report of Asteroseismology}, {LAURA} \citep{bruno2018}, searches the Pre-search Data Conditioning Simple Aperture Photometry, PDC-SAP FLUX light curves retrieved from Mikulski Archive for Space Telescopes, MAST or Kepler Asteroseismic Science Operations Center, KASOC databases and concatenates them in order to reduce noise in the data by using a high rate of oversample to maintain temporal stability by increasing the resolution of the points. After that, the code proceeds to do a  clean-up of excessive high values of frequencies thanks to a box-car type filter of 2$\mu$Hz. The time series extracted can then be treated by employing two different tools in order to characterize the periodicity of the star: Lomb-Scargle (LS) periodogram \citep{Scargle1982} and minimization of phase dispersion (PDM) \citep{stell1978}.
Figure \ref{pdm} shows the results obtained for the star HD~226808 (KIC~5307747) by using the method of minimization of phase dispersion and showing that the smallest phase occurs for a rotation period  of 122 days, in good agreement with the result obtained by Lomb-Scargle periodogram shown in  Figure~\ref{ls}. 
Our method based on the use of LS and PDM together, has given results consistent with the values found by  \cite{ceillier2017} who applied a different technique to a set of red giants, confirming the potential of our tool to determine the rotational period.  
The main packages of the code are presented in detail in Appendix \ref{appendixa}.

\begin{figure}
	\includegraphics[width=1.\columnwidth]{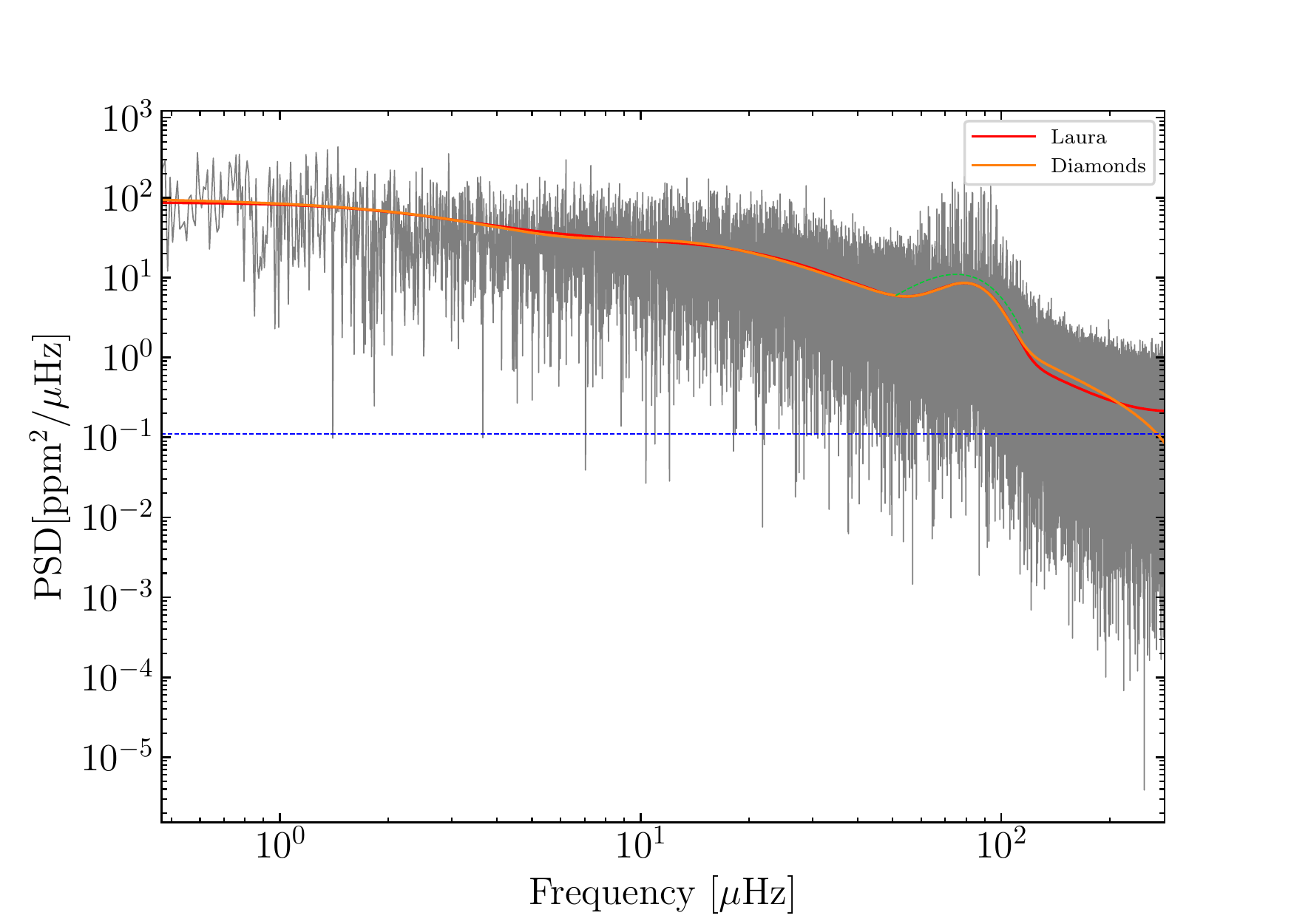}
	\caption{LAURA background model represented in comparison to DIAMONDS model by \citet{corsaro2015} for HD~226808 (KIC~5307747). White noise is shown by blue dashed line. Only subtle differences are present in the region close to the Nyquist frequency (283$\mu$Hz). The gaussian region highlighted in green is similarly analyzed in both.}
	\label{background}
\end{figure}

The same code has been implemented to perform the Fast Fourier Transform (FFT) of the time series to analyze the stellar oscillation power spectra and study 
excess of power at high frequency. The global approach in the seismic study that involves the adjustment of the entire spectrum has challenged us to compare with other already consolidated codes. From that comparison we finally  can conclude that LAURA is in good agreement with other ones using similar methods. For example here we show in Figure \ref{background}, the comparison of the background modeling methods between LAURA and DIAMONDS  \citep{corsaro2015} developed for the analysis of HD~226808. We found that the photometric granulation signatures due to intense convective activities on the surface 
present only subtle differences and do not affect the seismic analysis \citep{mathur2011}.

\begin{figure}
	
	\includegraphics[width=1.\columnwidth]{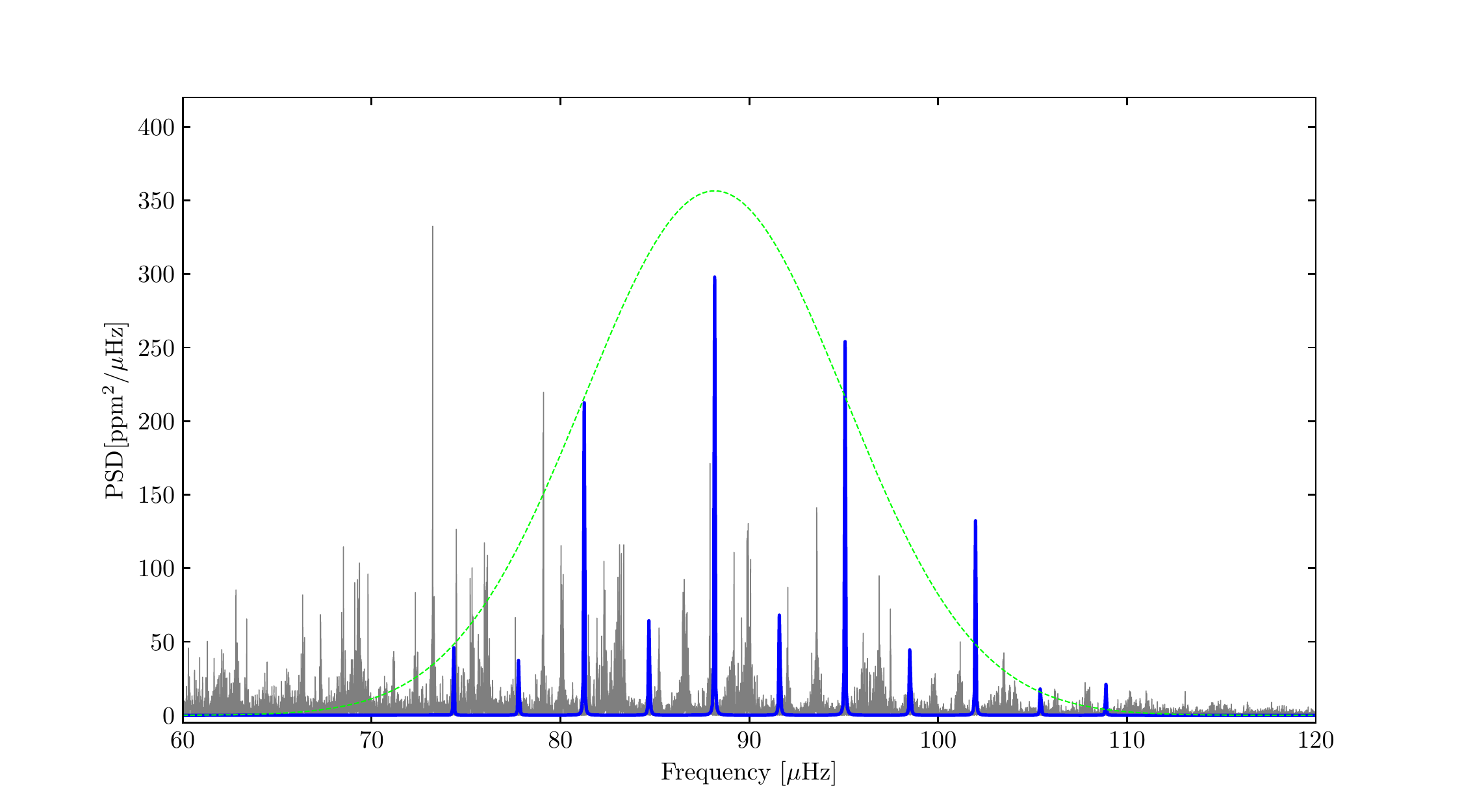}
	\caption{The gaussian shape of the forest region is shown in dashed green line. The central peak is the $\nu_{max}$ and the others in blue are separated by a periodicity of $\frac{\Delta \nu}{2}$.}
	\label{fig:Regiao}
\end{figure}

\begin{figure}
	\includegraphics[width=1.\columnwidth]{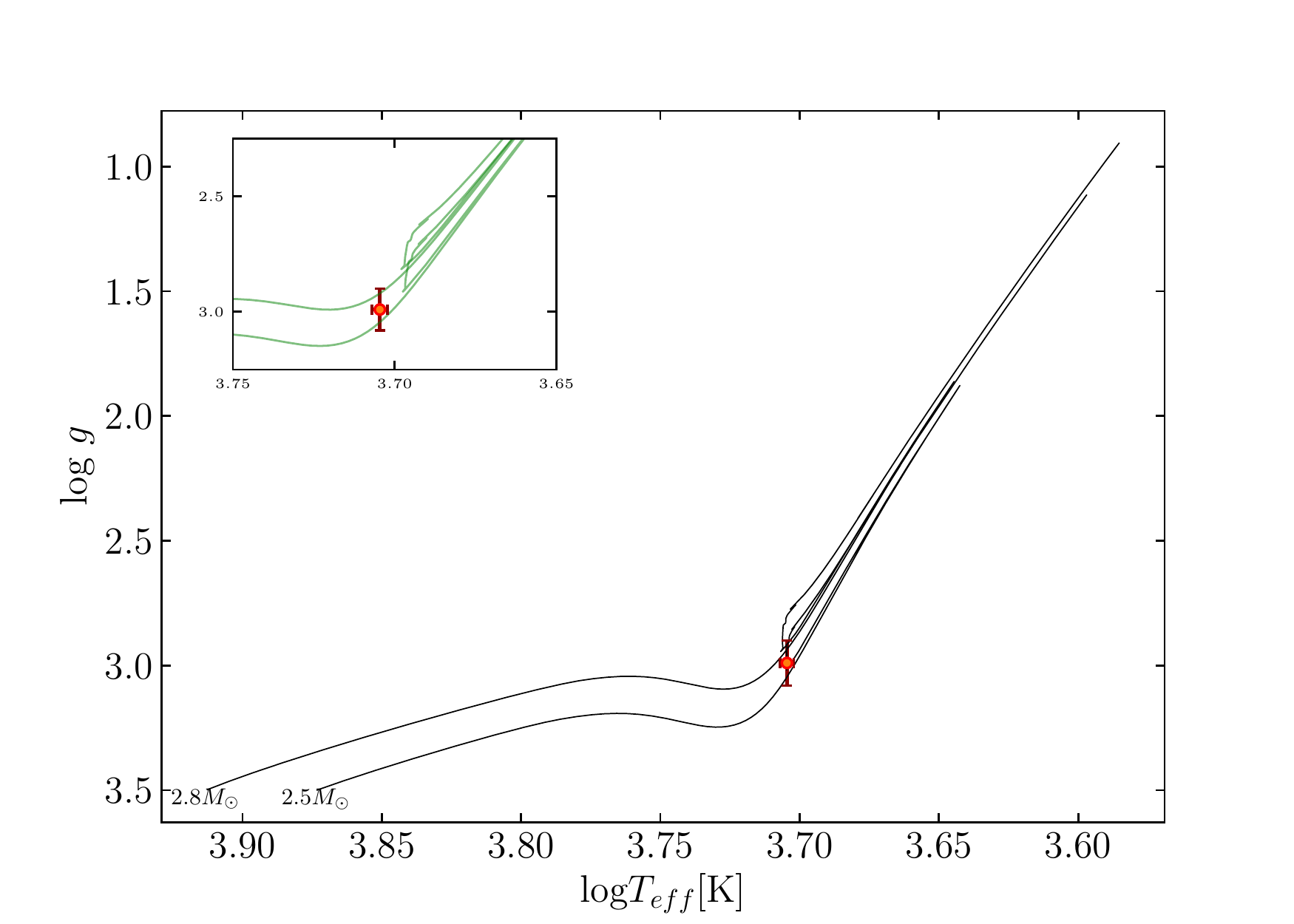}
	\caption{The red dot shows the location of HD~226808 in the $logg$ - $T_{eff}$ plane with two evolutionary tracks calculated for increasing values of the mass, while all the other parameters are fixed. The initial helium abundance is $Y_0=0.27$, the initial hydrogen abundance $X_0=0.71$. The values of the masses are given in the plot. The inset shows the location of the star with two evolutionary tracks obtained by assuming the solar metallicity. Using the metallicity determined by spectroscopy we see that it is a giant of the clump whereas it would be classified as an RGB in the solar metallicity.}
	\label{fig:DHR}
\end{figure}

\subsection{Results}

The power spectrum of the light curve for HD~226808 shows a clear excess of power, with a typical Gaussian distribution profile, in the range 60-120 $\mu$Hz due to the radial modes and some additional dipolar and quadrupolar modes, as it is shown in Figure~\ref{fig:Regiao}.
Based on  our method,  we determined the frequency at maximum oscillation power $\nu_{\rm max} = 87.91\pm1.9 \mu$Hz and the
so-called large frequency separation between modes with the same harmonic degree
$\Delta\nu = 6.998 \pm0.06 \mu$Hz.  The quoted uncertainties are
based on a local minimum optimization strategy called the Levenberg-Marquardt\footnote{https://lmfit.github.io/lmfit-py/intro.html} method in a specific package, allowing to estimate the following differences:
\begin{equation}
\chi^2 =  \sum_i^{N} \frac{[y^{\rm meas}_i - y_i^{\rm model}]^2}{\epsilon_i^2}
\end{equation}
where $y_i^{\rm meas}$, $y_i^{\rm model}$ and $\epsilon_i$ are the set of measured data, the model calculation and estimated uncertainty in the data, respectively. Using the universal method for red-giant oscillation pattern by \cite{mosser2011} and \cite{mosser2017} it was possible to identify pure-pressure modes and the identification of several dipole modes by grid-search allowed us to estimate the period spacing $\Delta \Pi = 296.85 \pm 6.53$s, which places this star on the secondary clump red-giant phase \citep{mosser2014}.
The asteroseismic fundamental parameters found here agree with the results published by \citet{mosser2012,mosser2014}, who
found for this star $\nu_{\rm max}=88.2 \mu$Hz,  and $\Delta \Pi=296.3\pm5.9$s, while their large separation, which is $\Delta \nu=6.88\pm0.05 \mu$Hz well agrees within the errors.
Table \ref{osc} lists the final set of frequencies for the detected individual modes together with their uncertainties.

\begin{table}
	\centering
	
	\begin{tabular}{ll}
		\hline
		\multicolumn{1}{c}{$l$} & \multicolumn{1}{c}{$\nu_{n,l}$}  \\ \hline
		0 &66.34916  $\pm$   0.3364\\ 
		0 &73.23633  $\pm$   0.0292\\ 
		0 &80.02904  $\pm$   0.2785\\ 
		0 &93.56723  $\pm$   0.0614\\
		1 &62.81494  $\pm$   0.0205\\ 
		1 &63.38166  $\pm$   0.0194\\
		1 &68.50571  $\pm$   0.0305\\
		1 &75.96745  $\pm$   0.0583\\
		1 &76.06977  $\pm$   0.0306\\
		1 &80.02891  $\pm$   0.0415\\
		1 &82.29577  $\pm$   0.0455\\
		1 &83.05140  $\pm$   0.0865\\ 
		1 &83.22456  $\pm$   0.0147\\ 
		1 &83.35049  $\pm$   0.0611\\ 
		1 &89.19081  $\pm$   0.0136\\ 
		1 &89.87559  $\pm$   0.0550\\ 
		1 &90.06449  $\pm$   0.0399\\ 
		1 &96.03074  $\pm$   0.0325\\
		2 &72.31541  $\pm$   0.3473\\ 
		2 &79.10025  $\pm$   0.0461\\ 
		2 &92.04025  $\pm$   0.1325\\ 
		3 &67.30144  $\pm$   0.0958\\ 
		3 &74.47983  $\pm$   0.0431\\ 
		3 &87.91571  $\pm$   0.0326\\ 
	\end{tabular}
	\caption{Frequencies and harmonic degree for  observed oscillation modes found  using LAURA script. Uncertainties are measured by Lorentzian $(l=0, 2, 3$) and sinc fits ($l=1$).}
	\label{osc}
\end{table}

\subsection{Fundamental parameters from asteroseismic scaling laws}
\label{laws}

The asteroseismic surface gravity,  stellar mass and radius for this star was  obtained
	from the observed $\Delta\nu$ and $\nu_{\mathrm{max}}$ together with the spectroscopic value 
	of $T_{\mathrm{eff}} =   5065^{+39}_{-29} \ $
	following the scaling relations calibrated on
	solar values as provided by \citep{KjeldsenBedding95,kallinger2010, belkacem2011}:
\begin{equation}
\frac{g_{ast}}{g_{\odot}}\simeq \Bigl(\frac{\nu_{max}}{\nu_{max \odot}}\Bigr)
\Bigl(\frac{T_{eff}}{T_{eff\odot}}\Bigr)^{1/2}
\label{eq_g}
\end{equation}
,
\begin{equation}
\frac{R_{ast}}{R_{\odot}}\simeq \Bigl(\frac{\nu_{max}}{\nu_{max \odot}}\Bigr)
\Bigl(\frac{\Delta\nu}{\Delta\nu_{\odot}}\Bigr)^{-2}
\Bigl(\frac{T_{eff}}{T_{eff\odot}}\Bigr)^{1/2}
\end{equation}
and
\begin{equation}
\frac{M_{ast}}{M_{\odot}}\simeq
\Bigl(\frac{\nu_{max}}{\nu_{max \odot}}\Bigr)^3
\Bigl(\frac{\Delta\nu}{\Delta\nu_{\odot}}\Bigr)^{-4}
\Bigl(\frac{T_{eff}}{T_{eff\odot}}\Bigr)^{3/2},
\end{equation}
where the values for the Sun are $\Delta\nu_{\odot}=134.9~\mu$Hz, $\nu_{max \odot}=3100~\mu$Hz and $T_{eff\odot}=5777~K$.
We obtained that $g_{ast}/g_{\odot}=0.027\pm0.003$, $M_{ast}/ M_{\odot} =2.59\pm0.24$
and $R_{ast}/ R_{\odot} = 9.86\pm0.613$. 


In comparison, the ratio $g_{spec}/g_{\odot}$ obtained in the spectroscopic analysis  a value of $\sim$0.036.
These values agree  within the errors  from asteroseismic radius and mass values computed  by 
\citet{mosser2012} and   \citet{takeda2015} and  based on different estimations of maximum power frequency.

\section{Comparison with evolutionary models}
Given the identified pulsation frequencies and the basic
atmospheric parameters, we faced the theoretical challenge
to interpret the observed oscillation modes by constructing stellar models which 
satisfy the spectroscopic and asteroseismic observational constraints. We assumed the effective temperature and surface gravity
as calculated in Section~2.1, respectively $T_{\rm eff}=5065^{+34}_{-29}$K, $\log g =2.99\pm0.08$~dex  and metallicity
$[Fe/H]=0.08\pm0.04$, as reported in (Table \ref{tab:01}).

We computed  theoretical evolutionary models representative of the present star by using the MESA evolutionary
code \citep{paxton2011}, in which we varied the stellar mass and the initial chemical composition in order to match the fundamental parameters available. 
\begin{figure}
	\includegraphics[width=1.\columnwidth]{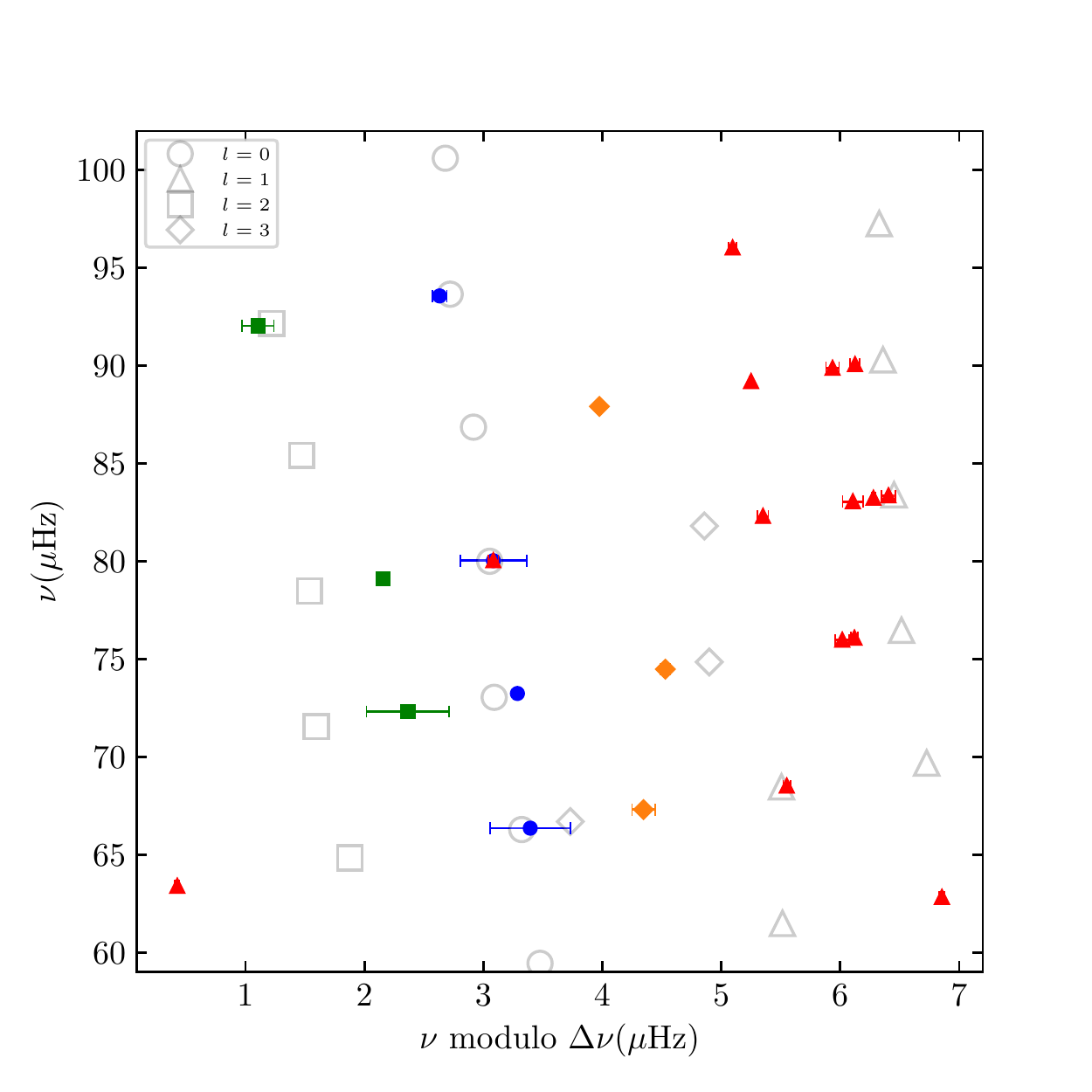}
	\caption{Echelle diagram for the  star HD~226808. The frequencies of observed  modes with  $l$ = 0, 1, 2, and 3 and their uncertainties are depicted in blue, red, green and orange, respectively.  Open symbols are the oscillation frequencies of our best-fitted model.}
	\label{fig:echelle}
\end{figure}
Figure~\ref{fig:DHR} shows two evolutionary tracks
obtained  with masses $2.5 M_{\odot}$  and  $2.8 M_{\odot}$ and fixed initial composition.  We used initial helium abundance of $Y_0=0.27$ and initial hydrogen abundance $X_0=0.71$ as input to the MESA and plotted in a  effective temperature-gravity evolutionary diagram. The location of HD~226808 indicated in Figure \ref{fig:DHR},   
which  identifies this star in the CHeB  phase of evolution (see Section.~1) with a mass between $2.5-2.8 M_{\odot}$.  
We point out that the location of evolutionary tracks in the diagram strongly depends on the metallicity of the model, since it influences the opacity and thus the depth of the convective envelope.  If a solar metallicity would have been assumed instead, as we can see in the insert in Figure~\ref{fig:DHR}, the star would have appeared to be in the RGB phase, burning hydrogen in a shell around the inert He core. As a result, a good determination of the measured metallicity is essential to achieve a correct determination of the evolutionary status of  stars located in this region of the HR diagram.  The $A(Li)$ also  confirm the evolved status of this star. 
\begin{figure}
	\includegraphics[width=1.\columnwidth]{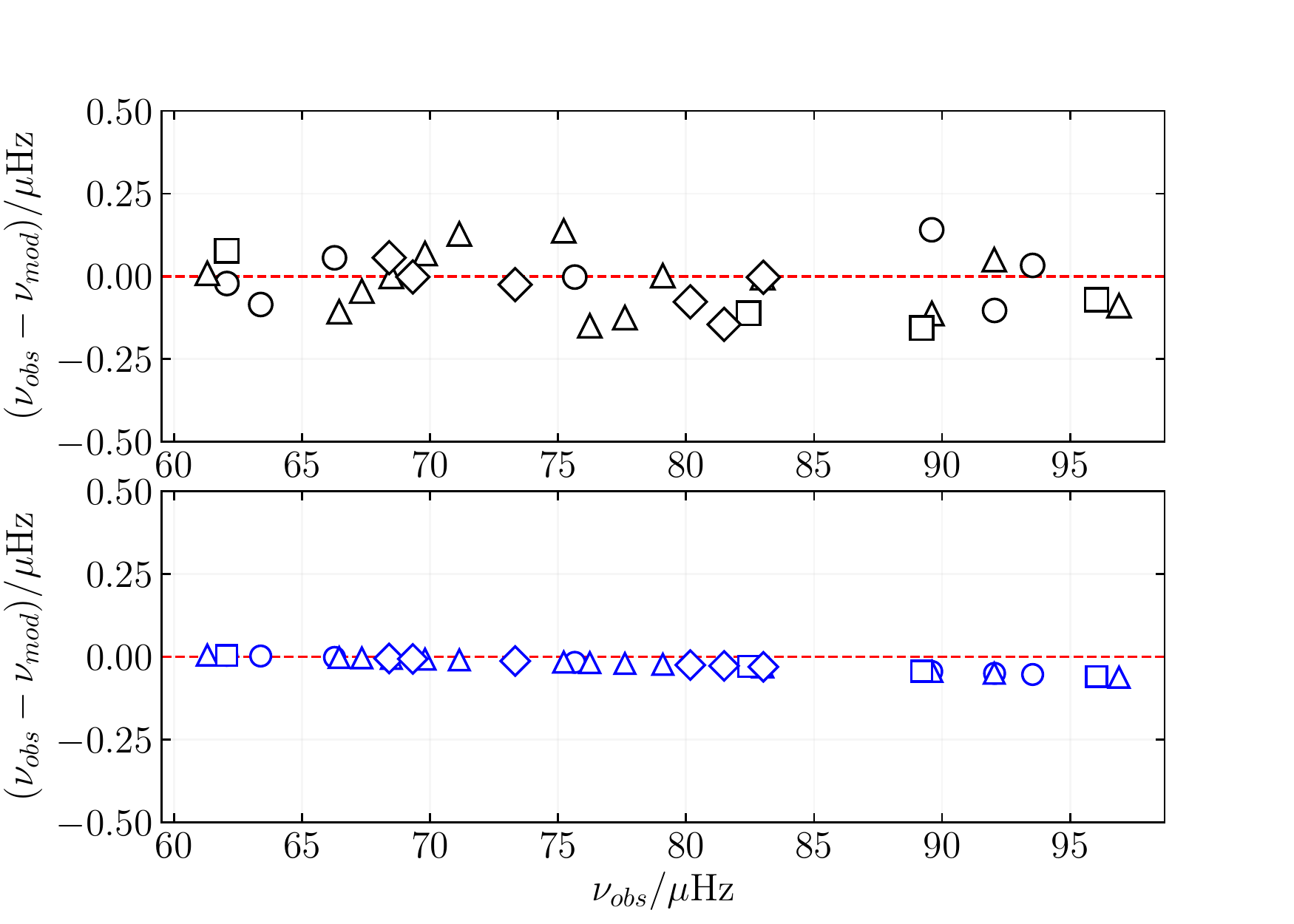}
	\caption{Top: difference between frequencies taken from LAURA without any surface correction  and those modeled using GYRE. Bottom: Application of surface corrections 
	 in order to reduce the differences shown on the top panel. Circles, triangles, squares and diamonds represent the modes $l = 0,1,2, 3$ , respectively.}
	\label{surface}
\end{figure}
In order to investigate the observed solar-like oscillations, we used the
GYRE package \citep{Townsend2013} to compute
adiabatic oscillation frequencies with degree $l=0,1,2, 3$ for some of the models
satisfying the spectroscopic constraints. In Figure~\ref{fig:echelle}, the open symbols represent the best output of the frequency model. 
Theoretical oscillation frequencies have been corrected by a near-surface effects term following the approach proposed by \citet{ball2014},  which has been proved to work much better for evolved stars than other prescriptions.
The adjust of the frequencies by a corrective term is a common procedure used for evolved stellar models to overcome the lack of a proper theory for the description of oscillations in the upper surface layers, where frequencies behaviors deviate from the adiabatic assumption.
In Figure~\ref{surface} shows the difference between modelled and observed frequencies for our best model with and without surface corrections, demonstrating the importance of such corrections.
Among our models we selected one that appeared to best fit the observed frequencies of HD~226808, and whose characteristics are given in   
Table~\ref{my-label}. This model is characterized by an age of  $0.69$\,Gyr, a mass $M=2.6\,  M_\odot$, a radius
$R=9.74{ R}_{\odot}$.  The value of the radius obtained by direct modelling of individual frequencies is now better constrained  and well agrees within the error bars obtained by using  scaling relation. 
Furthermore, the value of the mass derived from evolutionary track resulted to be in good agreement with respect to the seismic value, obtained by using the scaling relation and estimated in Section~\ref{laws}.  Large discrepancies by up to $ 50\%$ 
 between masses inferred from both methods was pointed out by \citet{miglio2012} and \citet{takeda2015}, who explored large sets of red-clump stars, and attributed to the overlapping of the evolutionary tracks of RG stars, with higher masses, and RC stars, with lower masses, at the same clump region of the HR diagram. This difference does not appear in the case of H-shell burning red giants.
\begin{table}
	\caption{Parameters of the best model using MESA for HD~226808 (KIC~5307747).}
	\centering
	
	\label{my-label}
	\begin{tabular}{lc}
		\multicolumn{1}{c}{Parameters} & Best model  \\ \hline
		$M/M_{\odot}$                          & 2.60   \\
		$R/R_{\odot}$                       & 9.74      \\
		$A(Li)$                             & 0.58    \\
		$\log T_{eff}$                       & 3.695     \\
		$\log g $ (cgs)                         & 2.856    \\
		Age (Gyr)                    & 0.69      \\
		$\Delta P$(s)                       & 221     \\
		\hline  
	\end{tabular}
\end{table}
Several tests considering interferometry \citep[e.g.,][]{Huber2012,white2013}, Hipparcos parallaxes \citep{aguirre2012}, eclipsing binaries \citep[e.g.,][]{frandsen2013, huber2015, gaulme2016} and open clusters \citep[e.g.,][]{miglio2012, miglio2016, stello2016},
have demonstrated that mass estimates from asteroseismic scaling relations are accurate to a few percent for main-sequence stars, while larger discrepancies have been identified for evolved stars.  Theoretical studies have suggested corrections to scaling
relations, for example by comparing the large frequency separation ($\Delta \nu$) calculated from individual frequencies
with models \citep[e.g.,][]{stello2009, white2011, 
	guggenberger2016, sharma2016} or an extension of the asymptotic relation \citep{mosser2013a}. For the moment, it remains clear that scaling relation corrections should mainly depend on $T_{eff}$ and evolutionary state.
However, it is yet unclear how these corrections
should be applied in the case of red clump stars \citep[e.g.,][]{miglio2012, sharma2016}.

\section{Conclusion}

In this study we perform an asteroseismic analysis of the red giant clump star HD 226808 (KIC 530774) based on its  high-precision space photometry   and  high-resolution ground spectroscopic observations.  The spectroscopic analysis was important to understand the metallicity of the star, which as discussed in Section 4, has an impact on the evolutionary state of the star.
Abundances obtained from the fundamental parameters solution, as for example $A(Li)$, produced good agreement between  observational and synthesized values.  We used the \textit{Kepler} photometric data as input to seismic diagnostic tools. We characterized  the internal structure and evolutionary status of this secondary clump red giant star, and studied  its frequencies,  especially since this 
stars it is one of the brightest  star of the \textit{Kepler} field. {\bf We have  shown} that accurate mass determination of RC stars with such a good agreement between  methods is possible. The metallicity effect on  oscillation modes for this class of stars  is still poorly understood. 
Future developments will be welcome to better understand this point, especially when comparing  giants  with 
very close evolutionary states. Finally, the presented method is robust enough to conduct seismic characterization of giants and and to obtain measurements 
of rotational period. Our tools are open source and ready for use.

\vspace{1cm}

\software {GYRE \citep{Townsend2013}, astropy (The Astropy Collaboration 2013, 2018), matplotlib \citep{Hunter2007},	
Numpy \citep{Walt2011}, Scipy \citep{scipy2001}, MOOG \citep{sneden1973}, ARES v2 \citep{sousa2007,  sousa2015}}

\section*{Acknowledgements}

We are grateful to the "National Council for Scientific and Technological Development" (CNPq, Brazil)
 for support.   JDN is supported by the  CNPq  Brazil PQ1 grant 310078/2015-6. PGB acknowledges the support of the MINECO under the fellowship program 'Juan de la Cierva incorporacion' (IJCI-2015-26034).
Based on observations obtained with the HERMES spectrograph on the Mercator Telescope, which is supported by the Research Foundation 
- Flanders (FWO), Belgium, the Research Council of KU Leuven, Belgium, the Fonds National de la Recherche Scientifique (F.R.S.-FNRS), 
Belgium, the Royal Observatory of Belgium, the Observatoire de Gen\'eve, Switzerland and the Th\"uringer Landessternwarte Tautenburg, Germany.
Funding for the \textit{Kepler} mission is provided by NASA's.



\begin{thebibliography}{99}
	
	\bibitem[Anders et al.(2016)]{Anders2016} Anders, F., Chiappini, C., Rodrigues, T.~S., et al.\ 2016, Astronomische Nachrichten, 337, 926 
	
	\bibitem[Asplund et al.(2009)]{2009ARA&A..47..481A} Asplund, M., Grevesse, N., Sauval, A.~J., et al.\ 2009, \araa, 47, 481
	
	\bibitem[Astropy Collaboration et al.(2013)]{2013A&A...558A..33A} Astropy Collaboration, Robitaille, T.~P., Tollerud, E.~J., et al.\ 2013, \aap, 558, A33
	
	
	\bibitem[Baglin et al.(2006)]{baglin}
	Baglin, A., Auvergne, M., Boisnard, L. et al. 2006, in 36th COSPAR Scientific Assembly Plenary Meeting, Vol. 36, Meeting abstract from the CDROM, 3749
	
	
	\bibitem[Ball \& Gizon(2014)]{ball2014} Ball, W.~H., \& Gizon, L.\ 2014, \aap, 568, A123 
	
	\bibitem[Barban et al.(2004)]{barban2004} Barban, C., De Ridder, J., Mazumdar, A., et al.\ 2004, SF2A-2004: Semaine de l'Astrophysique Francaise, 203 
	
	\bibitem[Beck et al.(2016)]{beck2016} Beck, P.~G., Allende Prieto, C., Van Reeth, T., et al.\ 2016, \aap, 589, A27 
	
	
	\bibitem[Bedding et al.(2011)]{bedding2011} Bedding, T.~R., Mosser, B., Huber, D., et al.\ 2011, \nat, 471, 608 
	
	\bibitem[Belkacem et al.(2011)]{belkacem2011}
	Belkacem, K., Goupil, M. J., Dupret, M. A., et al., 2011, A\&A, 530, 142
	
	\bibitem[Bensby et al.(2014)]{bensby2014} Bensby, T., Feltzing, S., \& Oey, M.~S.\ 2014, \aap, 562, A71 
	
	\bibitem[Borucki et al.(2009)]{borucki2009} 
	Borucki, W., Koch, D., Batalha, N., et al.\ 2009, Transiting Planets, 253, 289 
	
	\bibitem[Bovy et al.(2014)]{bovy2014} Bovy, J., Nidever, D.~L., Rix, H.-W., et al.\ 2014, \apj, 790, 127 
	
	\bibitem[Bressan et al.(2013)]{bressan2013} Bressan, A., Marigo, P., Girardi, L., Nanni, A., \& Rubele, S.\ 2013, European Physical Journal Web of Conferences, 43, 03001 
	
	\bibitem[Casagrande et al.(2014)]{casagrande2014} Casagrande, L., Silva Aguirre, V., Stello, D., et al.\ 2014, \apj, 787, 110 
	
	\bibitem[Castelli \& Kurucz(2004)]{castelli2004} Castelli, F., \& Kurucz, R.~L.\ 2004, arXiv:astro-ph/0405087 
	
	\bibitem[Cannon(1970)]{cannon1970} Cannon, R.~D.\ 1970, \mnras, 150, 111 
	
	
	\bibitem[Ceillier et al.(2017)]{ceillier2017} 
	Ceillier, T., Tayar, J., Mathur, S., et al.\ 2017, \aap, 605, A111 
	
	\bibitem[Chaplin \& Miglio(2013)]{chaplinEmiglio2013} 
	Chaplin, W.~J., \& Miglio, A.\ 2013, \araa, 51, 353 
	
	
	\bibitem[Christensen-Dalsgaard(2008)]{dalsgaard2008} 
	Christensen-Dalsgaard, J.\ 2008, \apss, 316, 113
	
	
	\bibitem[Corsaro et al.(2015)]{corsaro2015} 
	Corsaro, E., De Ridder, J., \& Garc{\'{\i}}a, R.~A.\ 2015, \aap, 578, A76 
	
	\bibitem[Cuypers(1983)]{Cuypers1983} 
	Cuypers, J.\ 1983, \aap, 127, 186 
	
	
	\bibitem[Cutri et al.(2003)]{cutri2003} 
	Cutri, R.~M., Skrutskie, M.~F., van Dyk, S., et al.\ 2003, VizieR Online Data Catalog, 2246,  
	
	\bibitem[De Ridder et al.(2006)]{deridder2006} De Ridder, J., Barban, C., Carrier, F., et al.\ 2006, \aap, 448, 689 
	
	\bibitem[De Moura \& De Almeida(2018)]{bruno2018} De Moura, B.~L., \& De Almeida, L.\ 2018, doi:10.5281/zenodo.1323217
	
	\bibitem[Di Mauro et al.(2018)]{dimauro2018} Di Mauro, M.~P., Ventura, R., Corsaro, E., et al.\ 2018, \apj, 862, 9
	
	\bibitem[Epstein et al.(2010)]{epstein2010} Epstein, C.~R., Johnson, J.~A., Dong, S., et al.\ 2010, \apj, 709, 447 
	
	\bibitem[Negri \& Vestri(2017)]{peakutils} Lucas Hermann Negri, L. H, \& Vestri, C. \ 2017. doi:10.5281/zenodo.887917	
	
	\bibitem[Faulkner \& Cannon(1973)]{faulkner1973} Faulkner, D.~J., \& Cannon, R.~D.\ 1973, \apj, 180, 435 
	
	\bibitem[Foreman-Mackey(2018)]{kplr} Foreman-Mackey, D.\ 2018, Astrophysics Source Code Library, ascl:1807.027 
	
	\bibitem[Frandsen et al.(2002)]{frandsen2002} Frandsen, S., Carrier, F., Aerts, C., et al.\ 2002, \aap, 394, L5 
	
	\bibitem[Frandsen et al.(2013)]{frandsen2013} Frandsen, S., Lehmann, H., Hekker, S., et al.\ 2013, \aap, 556, A138 
	
	
	\bibitem[Garc{\'{\i}}a et al.(2011)]{garcia2011} 
	Garc{\'{\i}}a, R.~A., Hekker, S., Stello, D., et al.\ 2011, \mnras, 414, L6 
	
	\bibitem[Gaulme et al.(2016)]{gaulme2016} Gaulme, P., McKeever, J., Jackiewicz, J., et al.\ 2016, \apj, 832, 121 
	
	
	\bibitem[Gilliland et al.(2010)]{gilliland2010} Gilliland, R.~L., Brown, T.~M., Christensen-Dalsgaard, J., et al.\ 2010, \pasp, 122, 131 
	
	\bibitem[Girardi et al.(1998)]{girardi1998} Girardi, L., Groenewegen, M.~A.~T., Weiss, A., \& Salaris, M.\ 1998, \mnras, 301, 149 
	
	
	\bibitem[Girardi(1999)]{girardi1999} Girardi, L.\ 1999, \mnras, 308, 818 
	
	\bibitem[Gontcharov(2017)]{gontcharov2017} Gontcharov, G.~A.\ 2017, Astronomy Letters, 43, 545 
	
	\bibitem[Guggenberger et al.(2016)]{guggenberger2016} Guggenberger, E., Hekker, S., Basu, S., \& Bellinger, E.\ 2016, \mnras, 460, 4277 
	
	
	
	\bibitem[Hawkins et al.(2017)]{hawkins2017} Hawkins, K., Ting, Y.-S., \& Rix, H.-W.\ 2018, \apj, 853, 20   
		
	\bibitem[Hekker et al.(2011)]{hekker2011} Hekker, S., Gilliland, R.~L., Elsworth, Y., et al.\ 2011, \mnras, 414, 2594 
	
	\bibitem[Hekker \& Christensen-Dalsgaard(2017)]{hekker2017} Hekker, S., \& Christensen-Dalsgaard, J.\ 2017, \aapr, 25, 1 
	
	
	\bibitem[Hinkle \& Wallace(2005)]{hinkle2005} Hinkle, K., \& Wallace, L.\ 2005, Cosmic Abundances as Records of Stellar Evolution and Nucleosynthesis, 336, 321 
	
	
	\bibitem[H{\o}g et al.(2000)]{hog2000} H{\o}g, E., Fabricius, C., Makarov, V.~V., et al.\ 2000, \aap, 355, L27 
	
	\bibitem[Huber et al.(2012)]{Huber2012} Huber, D., Ireland, M.~J., Bedding, T.~R., et al.\ 2012, \apj, 760, 32 
	
	\bibitem[Hunter (2007)]{Hunter2007} Hunter, J. D.\ 2007, Computing In Science \& Engineering, 9, 90 
		
	\bibitem[Huber et al.(2014)]{Huber2014} Huber, D., Silva Aguirre, V., Matthews, J.~M., et al.\ 2014, \apjs, 211, 2 
	
	\bibitem[Huber(2015)]{huber2015} Huber, D.\ 2015, Giants of Eclipse: The {$\zeta$} Aurigae Stars and Other Binary Systems, 408, 169 
	
	\bibitem[Jones et al.(2001)]{scipy2001} Jones, E, Oliphant, T., Peterson, P. et al, \ 2001,  Open source scientific tools for Python
	
	\bibitem[Kallinger et al.(2010)]{kallinger2010}
	Kallinger, T., Mosser, B., Hekker, S. et al., 2010, A\&A, 522, A1
	
	\bibitem[Kallinger et al.(2014)]{kallinger2014} Kallinger, T., De Ridder, J., Hekker, S., et al.\ 2014, \aap, 570, A41 
	\bibitem[Kjeldsen \& Bedding(1995)]{KjeldsenBedding95}
	Kjeldsen, H. \& Bedding, T., 1995, A\&A, 293, 87
	
	\bibitem[Kjeldsen et al.(2008)]{kjeldsen2008} Kjeldsen, H., Bedding, T.~R., \& Christensen-Dalsgaard, J.\ 2008, \apjl, 683, L175 
	
	{\bibitem[Bharat Kumar et al.(2018)]{kumar2018} Bharat Kumar, Y., Singh, R., Eswar Reddy, B., et al.\ 2018, \apjl, 858, L22}
	
	\bibitem[Lindegren et al.(2016)]{lindegren2016} Lindegren, L., Lammers, U., Bastian, U., et al.\ 2016, \aap, 595, A4 
	
	\bibitem[Mathur et al.(2010)]{mathur2010} Mathur, S., Garc{\'{\i}}a, R.~A., R{\'e}gulo, C., et al.\ 2010, \aap, 511, A46 
	
	\bibitem[Mathur et al.(2011)]{mathur2011} Mathur, S., Hekker, S., Trampedach, R., et al.\ 2011, \apj, 741, 119 
	
	\bibitem[Mel{\'e}ndez et al.(2012)]{mele2012} Mel{\'e}ndez, J., Bergemann, M., Cohen, J.~G., et al.\ 2012, \aap, 543, A29 
	
	
	\bibitem[Miglio et al.(2012)]{miglio2012} Miglio, A., Brogaard, K., Stello, D., et al.\ 2012, \mnras, 419, 2077 
	
	\bibitem[Miglio et al.(2016)]{miglio2016} Miglio, A., Chaplin, W.~J., Brogaard, K., et al.\ 2016, \mnras, 461, 760 
	
	
	\bibitem[Monroe et al.(2013)]{monroe2013} Monroe, T.~R., Mel{\'e}ndez, J., Ram{\'{\i}}rez, I., et al.\ 2013, \apjl, 774, L32 
	
	\bibitem[Mosser(2010)]{mosser2010} Mosser, B.\ 2010, Astronomische Nachrichten, 331, 944 
	
	
	\bibitem[Mosser et al.(2011)]{mosser2011} Mosser, B., Belkacem, K., Goupil, M.~J., et al.\ 2011, \aap, 525, L9 
	
	
	\bibitem[Mosser et al.(2012)]{mosser2012} Mosser, B., Goupil, M.~J., Belkacem, K., et al.\ 2012, \aap, 548, A10 
	
	\bibitem[Mosser et al.(2013)]{mosser2013a}
	Mosser, B., Michel, E., Belkacem, K., et al. 2013a, A\&A, 550, 126
	
	\bibitem[Mosser et al.(2014)]{mosser2014} Mosser, B., Benomar, O., Belkacem, K., et al.\ 2014, \aap, 572, L5 
	
	\bibitem[Mosser et al.(2017)]{mosser2017} Mosser, B., Pin{\c c}on, C., Belkacem, K., Takata, M., \& Vrard, M.\ 2017, \aap, 607, C2 
	
	\bibitem[Newville et al.(2018)]{lmfit} Newville, M., Otten, R., Nelson, et al.\ 2018, doi:10.5281/zenodo.1301254
	
	\bibitem[Paxton et al.(2011)]{paxton2011} Paxton, B., Bildsten, L., Dotter, A., et al.\ 2011, \apjs, 192, 3 
	
	\bibitem[Paxton et al.(2013)]{paxton2013} Paxton, B., Cantiello, M., Arras, P., et al.\ 2013, \apjs, 208, 4 
	
	\bibitem[Paxton et al.(2015)]{paxton2015} Paxton, B., Marchant, P., Schwab, J., et al.\ 2015, \apjs, 220, 15 
	
	\bibitem[Pinsonneault et al.(2014)]{pinsonneault2014} Pinsonneault, M.~H., Elsworth, Y., Epstein, C., et al.\ 2014, \apjs, 215, 19 
	
	\bibitem[Scargle(1982)]{Scargle1982} Scargle, J.~D.\ 1982, \apj, 263, 835 
	
	\bibitem[Sharma et al.(2016)]{sharma2016} Sharma, S., Stello, D., Bland-Hawthorn, J., Huber, D., \& Bedding, T.~R.\ 2016, \apj, 822, 15 
	
	\bibitem[Sneden(1973)]{sneden1973} Sneden, C.\ 1973, \apj, 184, 839 
	
	\bibitem[Sousa et al.(2007)]{sousa2007} Sousa, S.~G., Santos, N.~C., Israelian, G., Mayor, M., \& Monteiro, M.~J.~P.~F.~G.\ 2007, \aap, 469, 783 
	
	\bibitem[Sousa et al.(2015)]{sousa2015} Sousa, S.~G., Santos, N.~C., Adibekyan, V., Delgado-Mena, E., \& Israelian, G.\ 2015, \aap, 577, A67 
	
	
	\bibitem[Stanek et al.(1998)]{stanek1998} Stanek, K.~Z., Zaritsky, D., \& Harris, J.\ 1998, \apjl, 500, L141 
	
	\bibitem[Stellingwerf(1978)]{stell1978} Stellingwerf, R.~F.\ 1978, \apj, 224, 953 
	
	\bibitem[Stello et al.(2009)]{stello2009} Stello, D., Chaplin, W.~J., Basu, S., Elsworth, Y., \& Bedding, T.~R.\ 2009, \mnras, 400, L80 
	
	
	\bibitem[Stello et al.(2016)]{stello2016} Stello, D., Vanderburg, A., Casagrande, L., et al.\ 2016, \apj, 832, 133 
	
	\bibitem[Takeda \& Tajitsu(2015)]{takeda2015} Takeda, Y. \& Tajitsu, A. 2015 MNRAS 450,397
	
	\bibitem[Takeda et al.(2016)]{takeda2016} Takeda, Y., Tajitsu, A., Sato, B., et al.\ 2016, \mnras, 457, 4454 
	
	\bibitem[Tassoul(1980)]{tassoul1980} Tassoul, M.\ 1980, \apjs, 43, 469 
	
	\bibitem[Townsend \& Teitler(2013)]{Townsend2013} Townsend, R.~H.~D., \& Teitler, S.~A.\ 2013, \mnras, 435, 3406 
	
	
	\bibitem[Zahn(1992)]{zahn1992} Zahn, J.-P.\ 1992, \aap, 265, 115 
	
	\bibitem[Zechmeister \& K{\"u}rster(2009)]{Zechmeister2009} Zechmeister, M., \& K{\"u}rster, M.\ 2009, \aap, 496, 577 
	
	
	\bibitem[Ram{\'{\i}}rez \& Allende Prieto(2011)]{ramirez2011} Ram{\'{\i}}rez, I., \& Allende Prieto, C.\ 2011, \apj, 743, 135 
	
	\bibitem[Ram{\'{\i}}rez et al.(2014)]{ramirez2014} Ram{\'{\i}}rez, I., Bajkova, A.~T., Bobylev, V.~V., et al.\ 2014, \apj, 787, 154 
	
	\bibitem[{{Raskin}(2011)}]{RaskinPhD}
	{Raskin}, G. 2011, PhD thesis, Institute of Astronomy, Katholieke Universiteit
	Leuven, Belgium
	
	
	\bibitem[{{Raskin} {et~al.}(2011){Raskin}, {van Winckel}, {Hensberge},
		{Jorissen}, {Lehmann}, {Waelkens}, {Avila}, {de Cuyper}, {Degroote},
		{Dubosson}, {Dumortier}, {Fr{\'e}mat}, {Laux}, {Michaud}, {Morren}, {Perez
			Padilla}, {Pessemier}, {Prins}, {Smolders}, {van Eck}, \&
		{Winkler}}]{Raskin2011}
		
	{Raskin}, G., {van Winckel}, H., {Hensberge}, H., {et~al.} 2011, A\&A, 526, A69
	
	\bibitem[Silva Aguirre et al.(2012)]{aguirre2012} Silva Aguirre, V., Casagrande, L., Basu, S., et al.\ 2012, \apj, 757, 99 
	

	\bibitem[Walt et al.(2011)]{Walt2011}  Walt, S.~v., Colbert, S.~C., Varoquaux, G.\ 2011, Computing in Science \& Engineering, 13, 22 

	
	\bibitem[White et al.(2011)]{white2011} White, T.~R., Bedding, T.~R., Stello, D., et al.\ 2011, \apj, 743, 161 
	
	\bibitem[White et al.(2013)]{white2013} White, T.~R., Huber, D., Maestro, V., et al.\ 2013, \mnras, 433, 1262 
	
	
	
\end{thebibliography}


\appendix

\section{Master-Laura}
\label{appendixa}

\subsection{The algorithm}

LAURA (\textit{\textbf{L}ets  \textbf{A}nalysis,  \textbf{U}se and  \textbf{R}eport of 
	\textbf{A}steroseismology}) is written in Python 3 with 5
different packages to download, reduce and analyze raw light curves under the asteroseismology
theory. The code uses an object oriented approach and can be easily used even for non-experts in 
program language. All the additional python packages that we use are free and easy to install using the pip package. Our code is summarized in only 2 files: $LAURA\_ master.py$ and 
$LAURA\_ aux.py$. From the $LAURA\_ master.py$ file, the user can do all steps to download, 
reduce and analyze the light curve.

\subsection{Installation } \label{bozomath1}

In order to use our code, make sure you have python 3.x installed with the follow packages: Astropy \citep{2013A&A...558A..33A}, Scipy \citep{scipy2001}, kplr \citep{kplr}, peakutils \citep{peakutils}, and lmfit \citep{lmfit}. Installing the 
necessary packages in Ubuntu GNU/Linux are done using the follow:

\begin{lstlisting}
sudo apt-get install build-essential
sudo apt-get install python3
sudo apt-get install python3-pip
sudo pip install numpy
sudo pip install matplotlib
sudo pip install scipy
sudo pip install astropy
sudo pip install kplr
sudo pip install peakutils
sudo pip install lmfit

\end{lstlisting}

Download our source code from
\href{https://github.com/deMouradeAlmeida/LAURA}{GitHub} and add the path of where you want to install LAURA. In bash:

\begin{lstlisting}[language=Python]
export PYTHONPATH=/home/USER_NAME/LAURA-1.0.0/source:$PYTHONPATH
\end{lstlisting}

Add this command to your /.bashrc file or similar (/.cshrc, /.bashrc, /.profile). Go to your
LAURA folder and run the checkme.py file to verify if everything is in order. You should get 
a "good to go!" message. The installation in a OSX system can be done using Macports and more 
details can be found in the github guide.

\subsection{Packages} \label{bozomath}

LAURA runs from the box with 5 principal packages from the acquisition of the light curve to 
the Echelle diagram as follows:

\begin{lstlisting}[language=Python]
import LAURA_aux as aux

aux.TS_PS(ID,Qtin,Qtfin,cleaning,oversample,path = 'TEMP/')      
aux.FIT_PS(ID)                 
aux.LS(ID,kernerforca,Pbeg,Pend,PeriodMod,path='TEMP/')      
aux.OBS(ID)                    
aux.MOD(ID, largeSep,smallSep,size,Radial_Orders=5) 
aux.ECHELLE(ID, deltav)

print ('Done!')
\end{lstlisting}

All parameters configurations can be done using this file. The 
$TS\_PS$ package get the light curve, cleaning it and create the power spectrum using the time series analysis data. The input parameters are:

\begin{itemize}
	\item ID: Identification number of star
	\item Qtin: Inicial quarter for Kepler objects
	\item Qtfin: Final quarter for Kepler objects
	\item cleaning: Cleaning parameter for systematic errors
	\item Oversample: The resolution of the power spectrum
	\item Path: The folder in which the .fits files will be download
\end{itemize}

When running this package, LAURA will download all the .fits files from all the quarters you
select. If you already have this object in you computer, the code will not download again.
A power spectrum will be generated and all files will be store in a folder with the name of the
object.\\

\begin{figure}
	\centering
	\includegraphics[width=1.\columnwidth]{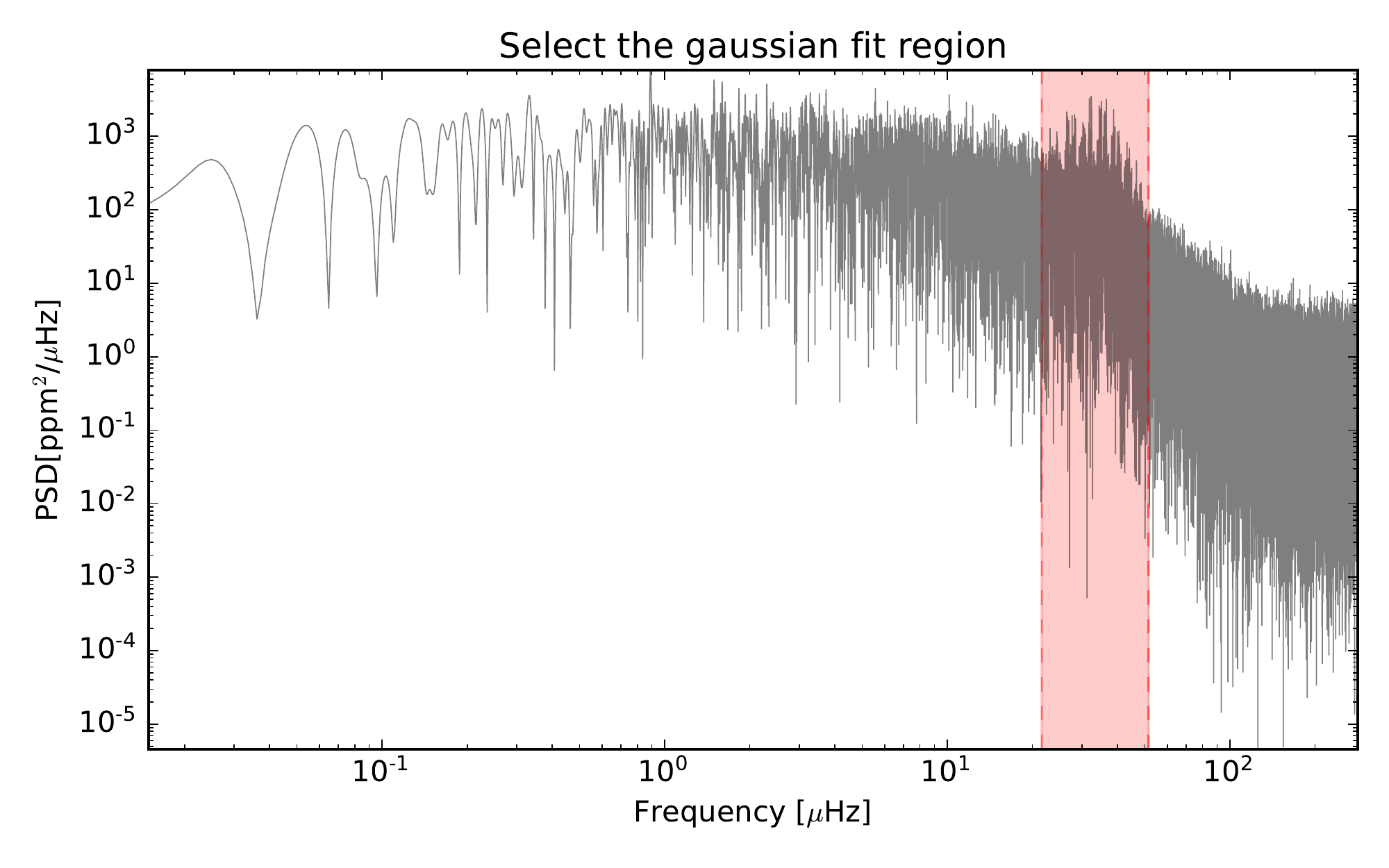}
	\caption{Power spectrum of KIC 5006817. The light-red region is the selection to fit
		a Gaussian fit.}
	\label{fitgauss}
\end{figure}

The $FIT\_PS$ package fits a background tendency to the power spectrum generated earlier using
only the ID of the object. The first plot will ask you to select the region to fit a Gaussian
curve as shown in the Figure \ref{fitgauss} following the equation \ref{gaussequation}

\begin{equation}
f(x) = \frac{A}{\sigma \sqrt{2 \pi}} e^{-\frac{(x-\mu)^2}{2 \sigma^2}}
\label{gaussequation}
\end{equation}

Where $A$ ins the amplitude, $\mu$ is the center and $\sigma$ is sigma.

Then, you will be ask to select two consecutive lorentzian fits by clicking the
regions of interest. Each click store a position to use as parameters of the 
following function:

\begin{equation}
f(x) = \frac{A}{\pi} \left ( \frac{\sigma}{(x-\mu)^2 + \sigma^2} \right )
\label{lorentzequation}
\end{equation}

All 3 models (one Gaussian and two Lorentzian) are then use to create a
background fit for the power spectrum as shown in the Figure \ref{background}.


The $LS$ package run a kernel to smooth out the light curve and run a periodgram
either using Lomb Scargle or Phase Dispersion Minimization (PDM). The input parameters are:

\begin{itemize}
	\item ID: Identification number of star
	\item Kernel: Strengh of the Gaussian 1D kernel
	\item Pin: Initial period of search
	\item Pend: Final period of search
	\item PeriodMod: Either LS (Lomb Scargle) or PDM (Phase Dispersion Minimization)
	\item Path: Where to save the data
\end{itemize}

This package will save all statistical results in a file named $kic\_periodgram.txt$.\\

The $OBS$ package return all the observed modes from the high frequency region
knowing only the ID of the object. First, the user will be asked to select the
high frequencies region from the power spectrum. Then it can be select 
the value of the threshold and the strength of the box kernel. The code then 
select all the peaks using a Lorentzian peak selector and saves all the output
data to the $\_data\_mods\_obs.txt$.\\

The $MOD$ package makes use of the universal pattern \cite{mosser2012} to model
the possible frequencies of the pure pressure modes of spherical degrees for a 
defined range of radial orders. It takes as input:

\begin{itemize}
	\item ID: Identification number of star
	\item LargeSep: large Frequency separation
	\item SmallSep: small Frequency separation
	\item Size: Size of the region which contains each mode
	\item Radial$\_$Orders: how many radial orders to calculate
\end{itemize}


The code plots the high frequency region from the observed data highlighting the regions  $l=0$ and $l=1$. This package will then save all model modes in a file named
$\_data\_mods\_model.txt$ that can be use to compute the Echelle diagram in the future.

For comparison, the example star mentioned here was fully analyzed in 22 minutes, 1 minute to download the light curves, around 20 minutes to process the raw LC into creating the power spectrum, and less than 1 second to automatically output the rotation period and asteroseismic parameters. Once we have the data download and the power spectrum done, the time consumed relies basically on the user interaction with the code, as LAURA is not fully automatic and needs the user to click and select regions in certain steps of the run. Thus, our code can't, for now, be used automatically to analyse  multiple stars without the user interaction.

\section{Line list}
\label{appendixb}

 \begin{table}
        \caption{\textbf{Full Line list used for HD~226808 (KIC\,530774)}}
        \centering
        \begin{tabular}{cccccc}
                \hline
                \multicolumn{1}{l}{$wavelength$} & \multicolumn{1}{c}{$species$} & \multicolumn{1}{c}{$\chi_{exc} $} & \multicolumn{1}{c}{$log (gf)$} & \multicolumn{1}{c}{EW (Sun)}  & \multicolumn{1}{c}{EW (HD 226808)} \\
                \multicolumn{1}{c}{$\mbox{\normalfont\AA} $} & \multicolumn{1}{c}{$   $} & \multicolumn{1}{c}{$eV$} & \multicolumn{1}{c}{$    $} & \multicolumn{1}{c}{\textbf{$\mbox{\normalfont\AA} $}}  & \multicolumn{1}{c}{\textbf{$\mbox{\normalfont\AA} $}} \\ \hline

\textbf{5295.3101} & \textbf{26.0}   & \textbf{4.42} & \textbf{-1.59} & \textbf{29.0}   & \textbf{50.9}  \\
\textbf{5379.5698} & \textbf{26.0}   & \textbf{3.69} & \textbf{-1.51} & \textbf{62.5} & \textbf{92.3}  \\
\textbf{5386.3301} & \textbf{26.0}   & \textbf{4.15} & \textbf{-1.67} & \textbf{32.6} & \textbf{59.5}  \\
\textbf{5441.3398} & \textbf{26.0}   & \textbf{4.31} & \textbf{-1.63} & \textbf{32.5} & \textbf{58.3}  \\
\textbf{5638.2598} & \textbf{26.0}   & \textbf{4.22} & \textbf{-0.77} & \textbf{80.0}   & \textbf{102.7} \\
\textbf{5679.0229} & \textbf{26.0}   & \textbf{4.65} & \textbf{-0.75} & \textbf{59.6} & \textbf{78.7}  \\
\textbf{5705.4639} & \textbf{26.0}   & \textbf{4.30}  & \textbf{-1.35} & \textbf{38.0}   & \textbf{62.1}  \\
\textbf{5731.7598} & \textbf{26.0}   & \textbf{4.26} & \textbf{-1.20}  & \textbf{57.7} & \textbf{83.7}  \\
\textbf{5778.4531} & \textbf{26.0}   & \textbf{2.59} & \textbf{-3.44} & \textbf{22.1} & \textbf{60.7}  \\
\textbf{5793.9141} & \textbf{26.0}   & \textbf{4.22} & \textbf{-1.62} & \textbf{34.2} & \textbf{61.2}  \\
\textbf{5855.0762} & \textbf{26.0}   & \textbf{4.61} & \textbf{-1.48} & \textbf{22.4} & \textbf{43.3}  \\
\textbf{5905.6699} & \textbf{26.0}   & \textbf{4.65} & \textbf{-0.69} & \textbf{58.6} & \textbf{78.0}    \\
\textbf{5927.7900} & \textbf{26.0}   & \textbf{4.65} & \textbf{-0.99} & \textbf{42.9} & \textbf{63.9}  \\
\textbf{5929.6802} & \textbf{26.0}   & \textbf{4.55} & \textbf{-1.31} & \textbf{40.0}   & \textbf{62.6}  \\
\textbf{6003.0098} & \textbf{26.0}   & \textbf{3.88} & \textbf{-1.06} & \textbf{84.0}   & \textbf{109.6} \\
\textbf{6027.0498} & \textbf{26.0}   & \textbf{4.08} & \textbf{-1.09} & \textbf{64.2} & \textbf{91.9}  \\
\textbf{6056.0000} & \textbf{26.0}   & \textbf{4.73} & \textbf{-0.40}  & \textbf{72.6} & \textbf{92.2}  \\
\textbf{6079.0098} & \textbf{26.0}   & \textbf{4.65} & \textbf{-1.02} & \textbf{45.6} & \textbf{68.2}  \\
\textbf{6093.6440} & \textbf{26.0}   & \textbf{4.61} & \textbf{-1.30}  & \textbf{30.9} & \textbf{55.4}  \\
\textbf{6096.6650} & \textbf{26.0}   & \textbf{3.98} & \textbf{-1.81} & \textbf{37.6} & \textbf{64.4}  \\
\textbf{6151.6182} & \textbf{26.0}   & \textbf{2.17} & \textbf{-3.28} & \textbf{49.8} & \textbf{93.0}    \\
\textbf{6165.3599} & \textbf{26.0}   & \textbf{4.14} & \textbf{-1.46} & \textbf{44.8} & \textbf{72.4}  \\
\textbf{6187.9902} & \textbf{26.0}   & \textbf{3.94} & \textbf{-1.62} & \textbf{47.6} & \textbf{77.3}  \\
\textbf{6240.6460} & \textbf{26.0}   & \textbf{2.22} & \textbf{-3.29} & \textbf{48.2} & \textbf{91.9}  \\
\textbf{6270.2251} & \textbf{26.0}   & \textbf{2.86} & \textbf{-2.54} & \textbf{52.4} & \textbf{92.2}  \\
\textbf{6703.5669} & \textbf{26.0}   & \textbf{2.76} & \textbf{-3.02} & \textbf{36.8} & \textbf{80.3}  \\
\textbf{6705.1021} & \textbf{26.0}   & \textbf{4.61} & \textbf{-0.98} & \textbf{46.4} & \textbf{73.5}  \\
\textbf{6713.7451} & \textbf{26.0}   & \textbf{4.79} & \textbf{-1.40}  & \textbf{21.2} & \textbf{39.6}  \\
\textbf{6726.6670} & \textbf{26.0}   & \textbf{4.61} & \textbf{-1.03} & \textbf{46.9} & \textbf{68.8}  \\
\textbf{6793.2588} & \textbf{26.0}   & \textbf{4.08} & \textbf{-2.33} & \textbf{12.8} & \textbf{32.4}  \\
\textbf{6810.2632} & \textbf{26.0}   & \textbf{4.61} & \textbf{-0.97} & \textbf{50.0}   & \textbf{73.7}  \\
\textbf{6828.5898} & \textbf{26.0}   & \textbf{4.64} & \textbf{-0.82} & \textbf{55.9} & \textbf{79.5}  \\
\textbf{6842.6899} & \textbf{26.0}   & \textbf{4.64} & \textbf{-1.22} & \textbf{39.1} & \textbf{62.4}  \\
\textbf{6843.6602} & \textbf{26.0}   & \textbf{4.55} & \textbf{-0.83} & \textbf{60.9} & \textbf{87.2}  \\
\textbf{6999.8799} & \textbf{26.0}   & \textbf{4.10}  & \textbf{-1.46} & \textbf{53.9} & \textbf{---}  \\
\textbf{7022.9502} & \textbf{26.0}   & \textbf{4.19} & \textbf{-1.15} & \textbf{64.5} & \textbf{---}  \\
\textbf{7132.9902} & \textbf{26.0}   & \textbf{4.08} & \textbf{-1.65} & \textbf{43.1} & \textbf{---}  \\
\textbf{4576.3330} & \textbf{26.1}   & \textbf{2.84} & \textbf{-2.95} & \textbf{64.6} & \textbf{111.2} \\
\textbf{4620.5132} & \textbf{26.1}   & \textbf{2.83} & \textbf{-3.21} & \textbf{50.4} & \textbf{78.8}  \\
\textbf{5234.6240} & \textbf{26.1}   & \textbf{3.22} & \textbf{-2.18} & \textbf{82.9} & \textbf{103.6} \\
\textbf{5264.8042} & \textbf{26.1}   & \textbf{3.23} & \textbf{-3.13} & \textbf{46.1} & \textbf{63.8}  \\
\textbf{5414.0718} & \textbf{26.1}   & \textbf{3.22} & \textbf{-3.58} & \textbf{27.3} & \textbf{45.2}  \\
\textbf{5425.2568} & \textbf{26.1}   & \textbf{3.20}  & \textbf{-3.22} & \textbf{41.9} & \textbf{60.5}  \\
\textbf{6369.4619} & \textbf{26.1}   & \textbf{2.89} & \textbf{-4.11} & \textbf{19.2} & \textbf{35.6}  \\
\textbf{6432.6758} & \textbf{26.1}   & \textbf{2.89} & \textbf{-3.57} & \textbf{41.3} & \textbf{62.7}  \\
\textbf{6516.0771} & \textbf{26.1}   & \textbf{2.89} & \textbf{-3.31} & \textbf{54.7} & \textbf{51.2}                                                      \\

                \hline
        \end{tabular}
        \label{linelist1}
        {\footnotesize \\
                \vspace{1mm} \textbf{$^{*}$  For species  we are using MOOG standard notation for atomic or molecular identification. \\
                For example, 26 represent Fe(26) while the 0 after the  decimal indicates neutral and 1 singly ionized.  \\
                 $\chi_{exc} $ is the line excitation potential  in electron volts (eV)}}
\end{table}


\label{lastpage}
\end{document}